\documentclass[12pt]{article}
\usepackage{epsfig,amsfonts,amsthm}
\usepackage{amsmath,amssymb}
\usepackage[normalem]{ulem}
\usepackage{color}
\usepackage{array}
\newcommand{\be}{\begin{equation}}
\newcommand{\ee}{\end{equation}}
\newcommand{\bea}{\begin{eqnarray}}
\newcommand{\eea}{\end{eqnarray}}

\newcommand{\doublet}[2]{ \left( \begin{array}{c}#1 \\ #2 \end{array}\right) }

\def\lsim{\mathrel{\rlap{\lower4pt\hbox{\hskip1pt$\sim$}}
    \raise1pt\hbox{$<$}}}         %less than or approx. symbol
\def\gsim{\mathrel{\rlap{\lower4pt\hbox{\hskip1pt$\sim$}}
    \raise1pt\hbox{$>$}}}         %greater than or approx. symbol
    
\usepackage{amsmath}
\usepackage{amsfonts}
\usepackage{amssymb}
\usepackage{subfig}
\usepackage{wrapfig}
\usepackage{graphicx}

\def\beq{\begin{equation}}
\def\eeq{\end{equation}}
\def\bea{\begin{eqnarray}}
\def\eea{\end{eqnarray}}

\def\<{\left\langle}
\def\>{\right\rangle}

\newcommand{\bt}{\begin{tabular}}
\newcommand{\et}{\end{tabular}}

\usepackage[
top    = 3.5cm,
bottom = 3.5cm,
left   = 3.5cm,
right  = 3.5cm]{geometry}

\allowdisplaybreaks[2]
\begin{document}
\bibliographystyle{OurBibTeX}

\title{\hfill ~\\[-30mm]
                  \textbf{Stabilizing the Higgs potential with a Z$'$}        
       }
\date{}
\author{\\[-5mm]
Stefano~Di~Chiara,\ 
Venus~Keus and\ 
Oleg~Lebedev \ \\ \\
  \emph{\small  Department of Physics and Helsinki Institute of Physics,}\\
  \emph{\small Gustaf H\"allstr\"omin katu 2a, 00014 University of Helsinki, Finland}\\
  \\[4mm]}
\maketitle

\vspace*{-0.250truecm}
\begin{abstract}
\noindent
{Current data point toward metastability of the electroweak vacuum within the Standard Model. We study the possibility of stabilizing the Higgs potential in U(1) extensions thereof. A generic Z$'$ boson improves stability of the scalar potential in two ways: it increases the Higgs self--coupling, due to a positive contribution to the beta--function of the latter, and it decreases the top quark Yukawa coupling, which again has a stabilizing effect. We determine the range of U(1) charges which leads to a stable electroweak vacuum. In certain classes of models, such stabilization is possible even if the Z$'$ does not couple to the Higgs and is due entirely to the reduction of the top Yukawa coupling. We also study the effect of the kinetic mixing between the extra U(1) and hypercharge
gauge fields. } 
\end{abstract}
 
\thispagestyle{empty}
\vfill
\newpage
\setcounter{page}{1}

\section{Introduction}

The current preferred values of the Higgs and top quark masses imply that the Higgs quartic coupling turns negative at some intermediate scale, signifying metastability of our electroweak vacuum \cite{Buttazzo:2013uya,Bezrukov:2012sa,Alekhin:2012py}. 
Although this is not problematic from the low energy perspective, it may lead to difficulties in reconciling the Standard Model (SM) with inflation, which entails large values for scalar fields in the Early Universe. Given the existence of a deep minimum at large Higgs values, the Universe is overwhelmingly likely to evolve to that catastrophic vacuum \cite{Espinosa:2007qp}.

This problem can be efficiently addressed by coupling the Higgs to the inflaton
\cite{Lebedev:2012sy}
thereby modifying the Higgs potential during inflation,
 yet other stabilizing mechanisms   are worth considering. The simplest possibility is to introduce a real \cite{Lebedev:2012zw,EliasMiro:2012ay} or complex \cite{Gonderinger:2012rd}  scalar which couples to the Higgs and makes the scalar potential convex. In this work, we consider the next simplest option: 
 introducing an extra U(1) symmetry \cite{Chao:2012mx,Chakrabortty:2013zja,Coriano:2014mpa}. The presence of an extra U(1) gauge 
 boson generally has a stabilizing effect on the potential due the positive contribution to the beta--function of the Higgs quartic coupling. We also find an additional positive  effect:   the top Yukawa coupling decreases with energy and therefore does not reduce the Higgs self--coupling as much as it does in the SM. In the framework of a generic U(1), we identify the main parameters responsible for the stabilization and consequently delineate our parameter space. 
 
To be as general as possible, we  avoid  working with specific charge assignments,
rather we single out important combinations thereof which make the main impact. 
We allow the ``hidden'' sector fields to be charged only under the 
extra U(1). When one assumes a generation-independent charge assignment,
anomaly--free models of this type can be parametrized in terms of
two parameters \cite{Appelquist:2002mw}. For our purposes this is not essential and our considerations apply to more general models. For example, the charges are allowed to be 
generation-dependent, if the corresponding Z$'$ is sufficiently heavy, in the 
range of a few TeV \cite{Langacker:2000ju}. Also, the extra U(1) may appear anomalous from the low energy perspective: its anomaly can be cancelled, though, by transforming a dilaton--like field, as in the Green--Schwarz mechanism \cite{Green:1984sg}. Such models are ubiquitous in realistic string constructions \cite{Buchmuller:2006ik}.
Therefore, we will not impose explicitly 
 the U(1) anomaly cancellation conditions and focus on a few charges, essential for our purposes , or combinations thereof. 

In what follows, we present the renormalization group equations for the relevant couplings and determine regions of parameter space consistent with stability of the Higgs potential, perturbativity, as well as the experimental constraints on a Z$'$.

\section{ Z$'$ framework}\label{U(1)-SM+S-potential}

We study extensions of the SM with the gauge group 
 $G_{SM} \times$ U(1)$'$ and additional SM singlet fields, both scalars and fermions. The SM fields generally carry charges under the extra U(1) as do the SM singlets. 
Allowing for a kinetic mixing between the U(1) and the hypercharge \cite{Holdom:1985ag},
 we take the Lagrangian  to be of the  form
\bea 
\mathcal{L}&=& 
\mathcal{L}^{\rm }_{SM} 
  + g_4 B'_\mu \sum_i  Q_i ~ \overline{\psi}_{\rm SM_{\it i}} \gamma^\mu ~ \psi_{\rm SM_{\it i}}  \\
&+& \sum_i D'_\mu S_i^* ~D'^\mu S_i  -\frac{1}{4}F'_{\mu\nu} F'^{\mu\nu}
- \frac{\epsilon}{2}F^Y_{\mu\nu} F'^{\mu\nu} \nonumber\\
&+& \sum_i \overline{\chi}_i (i\partial_\mu + g_4 Q'_i B'_\mu) \gamma^\mu \chi_i ~, \nonumber
\eea
where
\be
D'_\mu = \partial_\mu - ig_4 Q' B'_\mu ~,
\ee
and $B'_\mu$ and $g_4$ are the extra gauge field and its coupling.
Here $S_i$ and $\chi_i$ are the SM singlet scalars and fermions, respectively;
$Q_i$ and $Q_i'$ are the U(1) charges of the SM chiral fermions $\psi_{\rm SM_{\it i}}$ and the SM singlets, respectively.

We will further assume the scalar interaction between the Higgs field and the SM singlets as well as the singlet self-interaction to be small:
\be 
\Delta V =  \sum_{ij} \lambda_{s_i s_j} S_i^* S_i~ S^*_j S_j + \sum_i
\lambda_{hs_i} H^\dagger H~ S_i^* S_i~, \nonumber
\ee
with $\lambda_{s_i s_j},\lambda_{hs_i} \ll 1 $. Also we neglect effects  
of possible Yukawa couplings in the hidden sector.
These assumptions are not crucial, but they allow us to focus on the effects due to gauge interactions. The (stabilizing) effect of
the Higgs--portal couplings has been studied elsewhere (see for example \cite{Lebedev:2012zw}). 

The extra U(1) gets broken by one or more vacuum expectation values (VEVs)
of the singlets 
\begin{equation}
\langle S_i \rangle \not = 0 \;.
\end{equation}
We will not need the specifics of this breaking, except we will assume a single
scale at which the hidden sector activates and starts contributing to the RG equations for the SM couplings and $g_4$. This scale is associated with 
the gauge boson  (Z$'$) mass.
In general, one expects kinetic mixing between the
 Z$'$   and the Z. Phenomenology of Z$'$ models has been reviewed in \cite{Langacker:2008yv,Han:2013mra}.

\section{Sinopsis of  constraints}
In this section, we summarize the most important experimental constraints
on a $Z'$ as well as  theoretical constraints we impose on our models.

\begin{itemize}
\item
\textbf{Bounds from LEP}\\
LEP has set constraints on effective operators of the type
\begin{equation}
{g_{Z'}(i)~ g_{Z'}(j) \over m_{Z'}^2}~ \bar f_i \gamma^\mu f_i ~ \bar f_j \gamma_\mu f_j 
\end{equation}
for leptons $f_{i,j}$ of various chiralities with the Z$'$ couplings
 $ g_{Z'}(i), ~g_{Z'}(j)$. The strongest bound is set on
the vector-vector interactions \cite{Alcaraz:2006mx}:
\begin{equation}
{m_{Z'} \over \sqrt{ g_{Z'}(i) ~g_{Z'}(j)}} > 6.1 ~{\rm TeV} \;.
\end{equation}
Constraints on lepton--quark interactions are somewhat weaker and no useful bounds exist for  a leptophobic Z$'$.

\item
\textbf{Bounds from LHC}\\
{1. {\it General couplings}.} The most important bounds come from the CMS and ATLAS searches for 
dileptons with a large invariant mass, which result from
$q \bar q \to Z' \to \l^+ l^-$. We will use the CMS result 
\cite{Chatrchyan:2012oaa} as our benchmark constraint. 
For a sequential Z$'$, that is a Z$'$ with the couplings of the Z boson,
the bound is around 
\begin{equation}
m_{Z'_{\rm seq}} > 2.6 ~ {\rm TeV}\;.
\end{equation}
For a more general case, one needs to take into account both the difference in the couplings of the Z$'$ and the Z, and the reduction in the ``visible''
decay branching ratio due to the presence of new states \cite{Arcadi:2013qia}:
\begin{equation}
\sigma_{l^+ l^-} \to \left(  {g_{Z'} \over g_Z}  \right)^2 
{\rm BR_{vis}} ~ \sigma_{l^+ l^-} \;,
\end{equation}
where 
\begin{equation}
{\rm BR_{vis}} \simeq { \sum_i g_{Z'}(i)^2 \over 
 \sum_i g_{Z'}(i)^2 + \sum_{\chi_i}  g_{Z'}(\chi_i)^2}
\end{equation}
with $g_{Z'}(i)$ and $g_{Z'}(\chi_i)$ representing the couplings of the SM and extra fermions, respectively, into which the Z$'$ can 
decay\footnote{For simplicity we assume that the decay into the scalars is 
not allowed kinematically.}. 
As a result, the bound of $2.6$ TeV  can be relaxed and in some cases 
becomes as low as 500 GeV \cite{Lebedev:2014bba}. 
For our applications, we will typically take $m_{Z'}\sim$ 3 TeV to be 
on the safe side.
\\ \ \\
\noindent
{2. {\it Leptophobic} Z$'$.}  The bounds on a Z$'$ relax significantly
if it does not couple to the leptons. 
Taking the Z$'$ couplings to be of the electroweak size,
for $m_{Z'}> 2 m_t$, the typical
bounds are around 1 TeV. However, if  $m_{Z'}< 2 m_t$, the constraints
become very weak and the Z$'$ mass in the electroweak range is allowed.
A detailed analysis of this issue can be found in \cite{Chiang:2014yva}.

\item
\textbf{Stability of the Higgs potential}\\
This is a theoretical bound that we choose to impose,
\begin{equation}
\lambda_h >0
\end{equation}
at all scales up to the Planck scale. This ensures that the electroweak minimum
is stable. (Here we ignore possible complications associated with the SM singlet
directions in the scalar potential: we choose the couplings $\lambda_{h{s_i}}$
and $\lambda_{s_i}$ such that such issues do not arise.)

\item
\textbf{Perturbativity}\\
Assuming that the Z$'$ framework is valid up to the Planck scale, one 
must ensure perturbativity in this scale range.
In practice, we impose the condition
\begin{equation}
g_4^2~,\lambda_i  < 4 \pi 
\end{equation}
at the Planck scale, although the allowed parameter space is not sensitive 
to the exact value of the upper bound as long as it is ${\cal O}(1)$.

\end{itemize}

\section{RG evolution  of the couplings}
\subsection{The non-diagonal basis}

Following the method of \cite{Luo:2002iq}, we have derived the Renormalization Group Equations (RGEs) for the class of models in question. Given the substantial uncertainties associated with the top-quark mass, 
for our purposes it suffices
to use the 1--loop RGEs for most couplings.
In the beta function of $g_4$ however,
we take into account  the gauge two--loop contribution
proportional to $g_4^5$. This is because $g_4$ becomes substantial in certain
regions of parameter space we explore and the two--loop term may be relevant.
In general, neglecting the kinetic mixing
contributions, two-loop corrections can be incorporated using the numerical tool of \cite{Lyonnet:2013dna}.

Below we show the equations for the case of a single 
$S$ and a single $chiral$ fermion $\chi$. In the general case, one replaces
\footnote{In the gauge two loop contribution to $g_4$, one
replaces 
$Q_S^4 \rightarrow \sum_i Q_{S_i}^4 ~~,~~ Q_\chi^4 \rightarrow \sum_i Q_{\chi_i}^4 \;.$ } 
\begin{equation}
Q_S^2 \rightarrow \sum_i Q_{S_i}^2 ~~,~~ Q_\chi^2 \rightarrow \sum_i Q_{\chi_i}^2 \;.
\end{equation}
This is due to the fact that $S$ and $\chi$ do not couple to the SM gauge fields and their (additive) contributions are proportional to the charge squared. Therefore, in our discussion it is understood that $Q_S^2$ and $Q_\chi^2$ represent the sums over different species. 
Furthermore, in the RG equations for $g_4$ and $\epsilon$, we have also set the U(1) charges to be generation--independent. This can again be trivially generalized.

In the non--diagonal basis, i.e. the basis allowing for the kinetic mixing $\epsilon$, we find
\bea
16 \pi^2 \frac{d\lambda_h}{d t} &=& \lambda _h 
\left(-9 g_2^2 
-12 \zeta  g_4^2 Q_h^2
+12 \zeta  g_4 \epsilon  g_Y Q_h
-3 \zeta  g_Y^2 
+24 \lambda _h
+12 y_t^2\right)    \label{RG1} \\[1mm]
&&
+\frac{3}{8} \zeta ^2 g_Y^4
+\frac{3}{4} \zeta  g_2^2g_Y^2
+\frac{9 }{8}g_2^4
+\lambda _{hs}^2
-6 y_t^4 
+6 \zeta ^2 g_4^4 Q_h^4
-12 \zeta ^2 g_4^3 \epsilon  g_Y Q_h^3
\nonumber\\[1mm]
&&
+3 \zeta  g_4^2 g_2^2Q_h^2 
+3 \zeta ^2 g_4^2 g_Y^2 Q_h^2
+6 \zeta ^2 g_4^2 \epsilon ^2g_Y^2 Q_h^2
-3 \zeta ^2 g_4\epsilon  g_Y^3 Q_h
-3\zeta  g_4 g_2^2 \epsilon  g_Y Q_h ~, \nonumber\\[1mm]
16 \pi^2 \frac{d\lambda_{s}}{d t} &=& 
6 \zeta ^2 g_4^4 Q_S^4
-12 \zeta  g_4^2 Q_S^2 \lambda _s
+2 \lambda_{hs}^2
+20 \lambda _s^2 ~,\nonumber\\[1mm]
16 \pi^2 \frac{d\lambda_{hs}}{d t} &=& \lambda _{hs}
\left(-6 \zeta  g_4^2 Q_h^2 
+6 \zeta  g_4 \epsilon  g_Y Q_h 
-\frac{9}{2} g_2^2 
-6\zeta  g_4^2  Q_S^2
-\frac{3}{2} \zeta  g_Y^2
+12\lambda _h 
+8 \lambda _s
+6  y_t^2\right)\nonumber\\[1mm]
&& +4 \lambda _{hs}^2 
+12 \zeta ^2 g_4^4 Q_h^2 Q_S^2
-12 \zeta ^2 g_4^3\epsilon  g_Y Q_h Q_S^2 
+3 \zeta ^2 g_4^2 \epsilon ^2 g_Y^2 Q_S^2~,
\nonumber\\[1mm]
16 \pi^2 \frac{d y_{t}}{d t} &=& y_t
\left(-3 \zeta  g_4^2 Q_q^2 
+\zeta  g_4 \epsilon g_Y Q_q 
-3 \zeta  g_4^2 Q_t^2 
+4 \zeta  g_4 \epsilon  g_YQ_t 
-\frac{17}{12} \zeta  g_Y^2 
-\frac{9}{4} g_2^2 
-8g_3^2 
+\frac{9 }{2}y_t^2\right) ~,\nonumber\\[1mm]
16 \pi^2 \frac{d g_{4}}{d t} &=& g_4^3
\left(6 Q_b^2
+\frac{2}{3} Q_h^2
+4  Q_l^2
+12 Q_q^2
+\frac{1}{3}  Q_S^2
+6  Q_t^2
+2  Q_{\tau}^2
+\frac{2}{3}  Q_{\chi }^2 \right) \nonumber\\[1mm]
&&+\frac{1}{16\pi^2}\biggl(
 18 Q_b^4 
+8  Q_h^4
+12 Q_l^4
+36 Q_q^4
+4  Q_S^4
+18 Q_t^4
+6  Q_{\tau }^4
+2  Q_{\chi }^4
\biggr)g_4^5 ~,\nonumber\\[1mm]
16 \pi^2 \frac{d \epsilon}{d t} &=& \epsilon 
\left(\frac{2}{3}  Q_h^2
+4 Q_l^2
+12  Q_q^2
+6  Q_b^2
+\frac{1}{3}  Q_S^2
+6  Q_t^2
+2  Q_{\tau}^2
+{2 \over 3}  Q_{\chi }^2\right)g_4^2 
\nonumber\\[1mm]
&&
+\epsilon \frac{41}{6} g_Y^2
+ g_4 g_Y 
\left(4  Q_b
-\frac{2}{3} Q_h
+4 Q_l
-4  Q_q 
-8 Q_t
+4  Q_{\tau }\right)~,
 \nonumber
\eea
where $t=\ln (\mu/m_t)$ is the RG evolution variable and  $\zeta=1/(1-\epsilon^2)$. The SM gauge coupling RGEs are
\bea
\label{gauge-RGEs}
16 \pi^2 \frac{d g_{Y}}{d t} &=& \frac{41}{6} g_Y^3 ~,\\
16 \pi^2 \frac{d g_{2}}{d t} &=& -\frac{19}{6} g_2^3 ~,\nonumber\\
16 \pi^2 \frac{d g_{3}}{d t} &=& -7 g_3^3 ~,\nonumber
\eea
with the boundary values given in \cite{Lebedev:2012zw}.
The notation for the U(1) charges is straightforward: $Q_{t,b,\tau}$ are the charges 
for the right--handed fermions, while $Q_{q,l}$ are those for the left--handed fermions. As mentioned above, here we set them to be generation--independent.

The main new contribution to the running of $\lambda_h$ is due to the
positive terms proportional to $g_4^2$ and $g_4^4$ (unless $\epsilon$
is large). Also $y_t$ receives a new contribution with a definite sign:
the $g_4^2$--terms reduce the top Yukawa coupling. Both of these effects
increase $\lambda_h$ and tend to stabilize the Higgs potential. 

Let us note that $\lambda_{hs}$ and $\lambda_s$ are generated  by the RG 
evolution even if their initial values are zero. However, their numerical
impact on the evolution of $\lambda_h$ is not very significant in this case.

\subsection{Basis change}

It is often more convenient to work with the gauge fields which are orthogonal
at any energy scale. 
The kinetic mixing term $\frac{\epsilon}{2}F^Y_{\mu\nu} F'^{\mu\nu}$ in the Lagrangian can be rotated away
so that there is no mixing between $B^Y$ and $B'$ (see e.g. \cite{Datta:2013mta}). This is achieved
by the (RG scale--dependent)  transformation
\bea
F^Y \to \tilde{F^Y}- \frac{\epsilon\tilde{F'}}{\sqrt{1-\epsilon^2}} ~, \qquad F' \to \frac{\tilde{F'}}{\sqrt{1-\epsilon^2}} ~,
\eea
which leads to canonically normalized gauge fields.
The covariant derivative now contains the term:
\bea
  g_Y Y~ \tilde{B}^Y +\biggl(-\frac{\epsilon g_Y}{\sqrt{1-\epsilon^2}}  Y + \frac{g_4}{\sqrt{1-\epsilon^2}}Q \biggr)\tilde{B'}~,
 \label{D} 
\eea
which describes the relevant gauge interactions in the diagonal basis.
Defining the new coupling $\tilde g$ and redefining $g_4$ by
\bea
\tilde{g} = -\frac{\epsilon g_Y}{\sqrt{1-\epsilon^2}}~, \qquad  
 \frac{g_4}{\sqrt{1-\epsilon^2}} \to g_4~,
\eea
one can rewrite the RG equations in this diagonal basis. Note that no assumption on the smallness of $\epsilon$ has been made so far.

\subsection{The  diagonal basis}

In terms of the redefined couplings, the RG equations read:
\bea
16 \pi^2 \frac{d\lambda_h}{d t} &=& 
-6 y_t^4
-3 \biggl(
 g_1^2 
+3g_2^2 
+\tilde{g}^2
-8 \lambda_h
-4 y_t^2 \biggr) \lambda_h
\nonumber\\[1mm]
&&+\frac{3}{4} g_1^2\tilde{g}^2
+\frac{3}{4} g_2^2 \tilde{g}^2 
+\frac{3}{8}  \tilde{g}^4
+\frac{3 }{8}g_1^4
+\frac{9 }{8}g_2^4
+\frac{3}{4}g_1^2 g_2^2
+\lambda_{hs}^2\nonumber\\[1mm]
&& +3Q_h  \biggl(
g_1^2 
+g_2^2 
+\tilde{g}^2 
-4\lambda _h 
+4Q_h^2 g_4^2 \biggr)g_4\tilde{g} 
\nonumber\\[1mm]
&&
+ 3Q_h^2 \biggl(
 g_1^2 
+g_2^2
+3\tilde{g}^2 
-4 \lambda _h 
+2 Q_h^2 g_4^2 \biggr)g_4^2  ~,\nonumber\\[1mm]
16 \pi^2 \frac{d\lambda_{s}}{d t} &=& 
-12 g_4^2 Q_S^2 \lambda _s +6 g_4^4 Q_S^4+2 \lambda
   _{hs}^2+20 \lambda _s^2 ~,\nonumber\\[1mm]
16 \pi^2 \frac{d\lambda_{hs}}{d t} &=& 
\biggl( 6 y_t^2
+8 \lambda_s
+4 \lambda_{hs}
+12 \lambda_h 
-\frac{3}{2}\tilde{g}^2
-\frac{3}{2} g_1^2
-\frac{9}{2} g_2^2 \biggr)\lambda_{hs}  \nonumber\\[1mm]
&& -6 g_4 \biggl( 
\tilde{g} Q_h 
+ g_4 Q_h^2
+ g_4Q_S^2
\biggr)\lambda_{hs} 
+3g_4^2 \biggl(
\tilde{g}^2 
+4 g_4 \tilde{g} Q_h
+4 g_4^2 Q_h^2
\biggr) Q_S^2~,
\nonumber\\[1mm]
16 \pi^2 \frac{d g_{4}}{d t} &=&
\frac{41}{6}g_4 \tilde{g}^2
+\biggl(-4  Q_b
+\frac{2}{3}Q_h
-4 Q_l
+4  Q_q
+8 Q_t 
-4 Q_{\tau} \biggr) g_4^2 \tilde{g} \nonumber\\[1mm]
&&+\biggl(\overbrace{
6 Q_b^2 
+\frac{2}{3}  Q_h^2
+4 Q_l^2
+12Q_q^2
+\frac{1}{3}  Q_S^2
+6 Q_t^2
+2 Q_{\tau }^2
+{2\over 3} Q_{\chi }^2}^{Q_4^2}  \biggr)g_4^3\nonumber\\[1mm]
&&+\frac{1}{16\pi^2}\biggl(
 18 Q_b^4 
+8  Q_h^4
+12 Q_l^4
+36 Q_q^4
+4  Q_S^4
+18 Q_t^4
+6  Q_{\tau }^4
+2  Q_{\chi }^4
\biggr)g_4^5 ~,\nonumber\\[1mm]
16 \pi^2 \frac{d \tilde{g}}{d t} &=&
\frac{41}{6} \tilde{g}^3
 + \biggl(-4Q_b 
+\frac{2}{3} Q_h
-4 Q_l
+4  Q_q
+8 Q_t
-4  Q_{\tau}\biggr)g_4\tilde{g}^2 \nonumber\\[1mm]
&&+\frac{41}{3} g_1^2\tilde{g}
+\biggl(6 Q_b^2 
+\frac{2}{3}  Q_h^2
+4  Q_l^2
+12  Q_q^2
+\frac{1}{3} Q_S^2
+6  Q_t^2
+2 Q_{\tau }^2
+{2\over 3}  Q_{\chi }^2 \biggr)g_4^2 \tilde{g} \nonumber\\[1mm]
&&+ \biggl(-4 Q_b 
+\frac{2}{3}Q_h
-4 Q_l
+4  Q_q 
+8 Q_t
-4 Q_{\tau} \biggr) g_1^2 g_4 ~,\nonumber\\[1mm]
&&\nonumber\\
%%%
%
16 \pi^2 \frac{d y_{t}}{d t} &=& 
y_t\biggl(
- g_4 \tilde{g}( 4Q_t+Q_q )
-3 g_4^2  (Q_q^2+Q_t^2)
-\frac{17}{12} g_1^2
-\frac{9 }{4}g_2^2
-8g_3^2
-\frac{17}{12}\tilde{g}^2
+\frac{9}{2} y_t^2 
\biggr) ~, \nonumber
\eea
with the SM gauge coupling RGEs being the same as in Eq.(\ref{gauge-RGEs}).
This result agrees with known special cases such as U(1)$_{B-L}$ \cite{Basso:2010jm}.
We will mostly use these RG equations in our numerical analysis.

In the beta function of $g_4$,  the leading term for
small Z-Z$'$ mixing is associated with the combination of charges
which we call $Q_4^2$:
\begin{equation}
Q_4^2 \equiv 6 Q_b^2 
+\frac{2}{3}  Q_h^2
+4 Q_l^2
+12Q_q^2
+\frac{1}{3}  Q_S^2
+6 Q_t^2
+2 Q_{\tau }^2
+{2\over 3} Q_{\chi }^2 ~. \label{Q4}
\end{equation}
Therefore, most of the individual charges do not matter for our analysis as long as $Q_4$ remains the same. The other two important quantities
 are $Q_h$ and $Q_{t,q}$ for the third generation. The Higgs charge
 appears explicitly in the beta function for $\lambda_h$. Again, for small
 Z-Z$'$ mixing, it only contributes as $Q_h^2$. Similarly, the beta function of 
 the top Yukawa coupling depends on $Q_{q,t}^2$ at leading order.
 Since  $Q_q$ for the third generation and $Q_t$ are related by U(1) invariance
 of the Yukawa interaction, the essential parameters for our study are
 \begin{equation}
 Q_h^2 ~,~Q_{q_3}^2~,~Q_4^2~.
 \end{equation} 
Although we have imposed generation-independent charges in the above equations,
it is straightforward to adapt the RGE's to a non--universal case. Clearly, what matters for our purposes is the charge assignment for the top quark.

\section{Stabilizing the Higgs potential via a U(1)$'$}

The presence of an extra U(1) modifies the Higgs self--coupling $\lambda_h$
at energies above the Z$'$ mass scale. For small Z-Z$'$ mixing, the effect
is always positive. This is because the new contributions to the beta functions
increase $\lambda_h$ at high energies both through a direct one loop contribution to $\beta_{\lambda_h}$ and by decreasing the top Yukawa coupling:
\begin{eqnarray}
\Delta \beta_{\lambda_h} &\propto& g_4^2 Q_h^2  + c~ g_4^4 Q_h^4~, \nonumber \\
 \Delta \beta_{y_t} &\propto & -g_4^2 (Q_q^2 + Q_t^2)~.
 \label{delta}
\end{eqnarray}
(Here the proportionality coefficients are positive at $m_t$.)
This is a general feature of Z$'$ models. Therefore, additional U(1) symmetries
tend to stabilize the Higgs potential. In what follows, we study the specifics of this effect.

Note also that the Higgs--singlet coupling has a stabilizing effect too:
its contribution to $\beta_{\lambda_h}$ is positive and proportional to 
$\lambda_{hs}^2$. However, such a coupling is not specific to U(1)$'$
models and we choose to minimize this effect by setting $\lambda_{hs}\ll 1$,
$\lambda_{s}\ll 1$ at the electroweak scale.

In order for the SM Yukawa couplings to be allowed, the U(1) charges 
must satisfy the constraints
\be 
\label{gauge-invariance}
Q_t=Q_q +Q_h, \quad Q_b=Q_q-Q_h, \quad Q_\tau =Q_l-Q_h~.
\ee
Eliminating the charges of right--handed fermions, $Q_4$ takes the form
\be 
Q_4^2 = \frac{44}{3}Q_h^2 + 24Q_q^2 +\frac{1}{3} Q_S^2 
+{2\over 3} Q_{\chi}^2 +6 Q_l^2 -4Q_hQ_l ~. \label{Q4-2}
\ee
This combination of charges is responsible for most of the running of $g_4$, 
which in turn affects $\lambda_h$ as shown in Eq.~(\ref{delta}).  

Apart from their contribution to $Q_4$, the remaining charges $Q_S$, $Q_l$, and $Q_\chi$ have very little effect on the running of $\lambda_h$. They enter
the kinetic mixing and the 2--loop contributions, whose effect is clearly
subdominant. $Q_S$ also appears in the beta--functions of 
$\lambda_s$ and $\lambda_{hs}$. Due to our boundary conditions
$\lambda_{hs}\ll 1$, $\lambda_{s}\ll 1$, these couplings do not make a significant impact on $\lambda_h$. Therefore, $Q_S$ affects stability of the Higgs potential almost entirely through $Q_4$. 

Also, the quark charges for the first two generations $Q_{q_{1,2}}$ 
are not relevant for our analysis. For clarity however,
we make the universality assumption
\begin{equation}
Q_q = Q_{q_{3}} =Q_{q_{1,2}} ~.
\end{equation}

In what follows, we study the numerical impact of $Q_h, Q_q$ and $Q_4$.

\subsection{Generic Z$'$}

A generic Z$'$ is subject to the strong LEP and LHC constraints of Section 3. For
the Z$'$ coupling of electroweak size, we choose $M_{Z'}=3$ TeV as the reference point. That means we apply the RG equations of Section 4 above this
scale, whereas below 3 TeV the running is SM--like. We choose three representative values $g_4=0.1,~0.2,~0.3$. For larger $g_4$, the couplings  become
non--perturbative below the Planck scale (for order one charges).
In this subsection, we set the kinetic mixing to zero at $M_{Z'}$, that is
$\tilde g (M_{Z'})=0$, and focus entirely on the effect 
of $g_4$.

\begin{figure}[h!]
\centering
\includegraphics[width=0.42\linewidth]{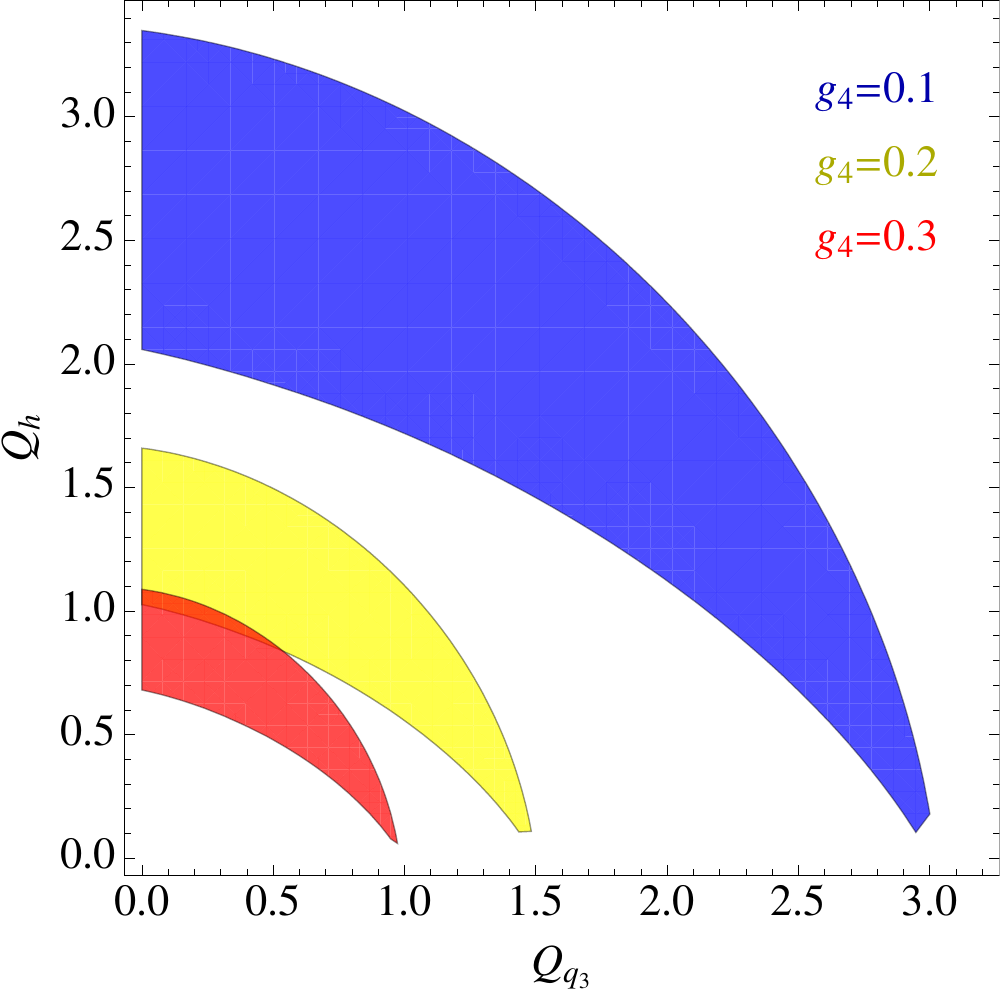}
\includegraphics[width=0.42\linewidth]{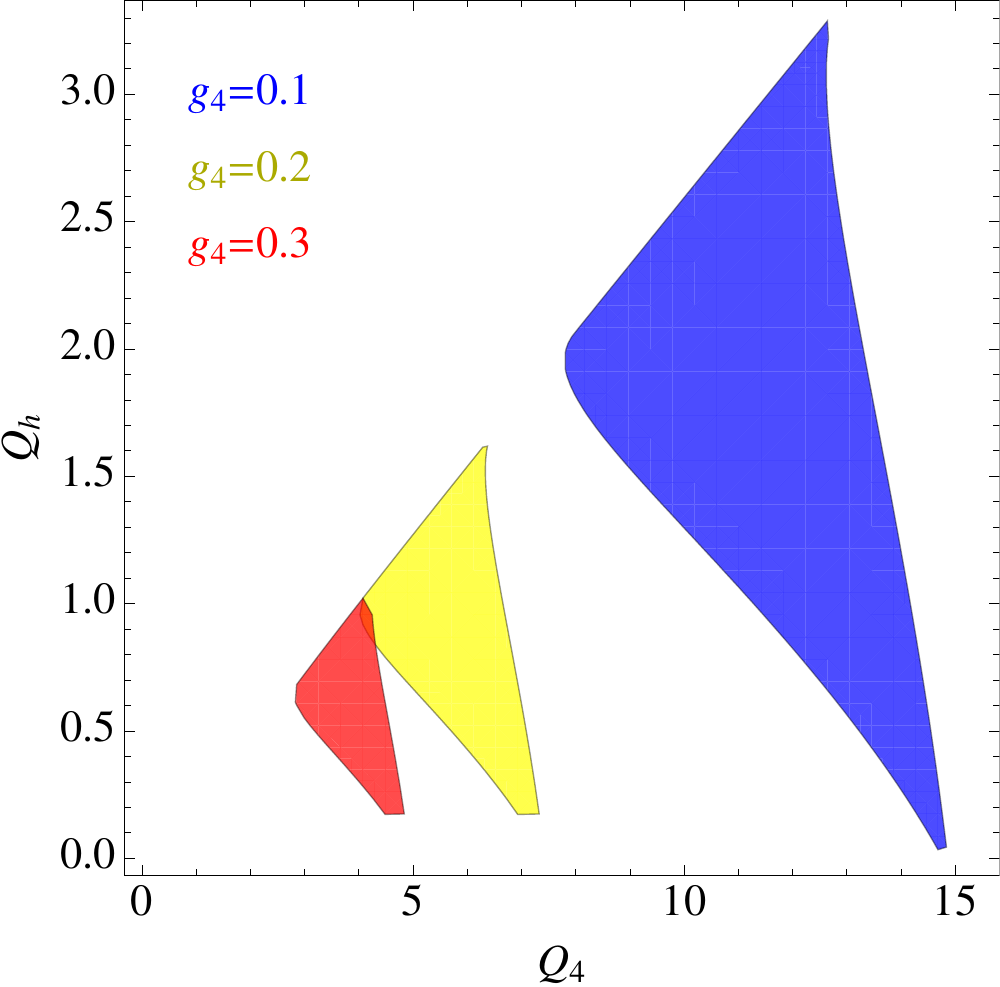}
\caption{Charge assignments consistent with Higgs potential stability and perturbativity for  $g_4=0.1$ (blue region), $g_4=0.2$ (yellow region) and $g_4=0.3$ (red region) at $M_{Z'}=3$ TeV.}
\label{region-plot-general}
\end{figure}

Figure~\ref{region-plot-general} shows regions of parameter space allowed by stability of the Higgs potential and perturbativity (shaded regions). As 3 charges play the most important role, we display our results in two planes: 
$Q_{q_3}-Q_h$ and $Q_4-Q_h$. To fix $Q_4$, we set 
$Q_l=Q_\chi=0,~ Q_S=2$, although, as explained above, the results do not depend on this choice. 

The shape of the allowed regions can be easily understood.
The lower bound on charges is dictated by stability of the Higgs potential, whereas the upper bound is imposed by perturbativity. The allowed regions shrink as $g_4$ increases since the couplings reach non--perturbative values sooner. The $Q_{q_3}-Q_h$ panel shows that the stabilization is possible for non--zero Higgs charges only, although such charges can be as small as
$10^{-1}$. As $Q_{q_3}$ increases, values of the allowed  Higgs charges decrease since
smaller charges are sufficient to stabilize the Higgs potential on one hand
and smaller charges are compatible with perturbativity on the other hand.  
Above a certain critical $Q_{q_3}$, no charge assignment leads to a 
perturbative result.

In the $Q_4-Q_h$ plane, we vary $Q_4$ by changing $Q_{q_3}$ while keeping
the rest of the charges intact. 
The upper bound on $Q_h$ at a given $Q_4$
is dictated by positivity of $Q_i^2$ in   Eq.~\ref{Q4} (or Eq.~\ref{Q4-2}).
Large values of $Q_4 $ violate perturbativity, while its low values for a 
fixed $Q_h$ would fail to stabilize the potential.

%%%%%%%%%%%%%%%%%%%%%%%%%%%%%%%%%%%%%%%%%%%%%%%%%%%%%%%%%%%%%%%%%%%%%%%%%%%%%%

\subsection{Leptophobic $Z'$}
\begin{figure}[h!]
\centering
\includegraphics[width=0.42\linewidth]{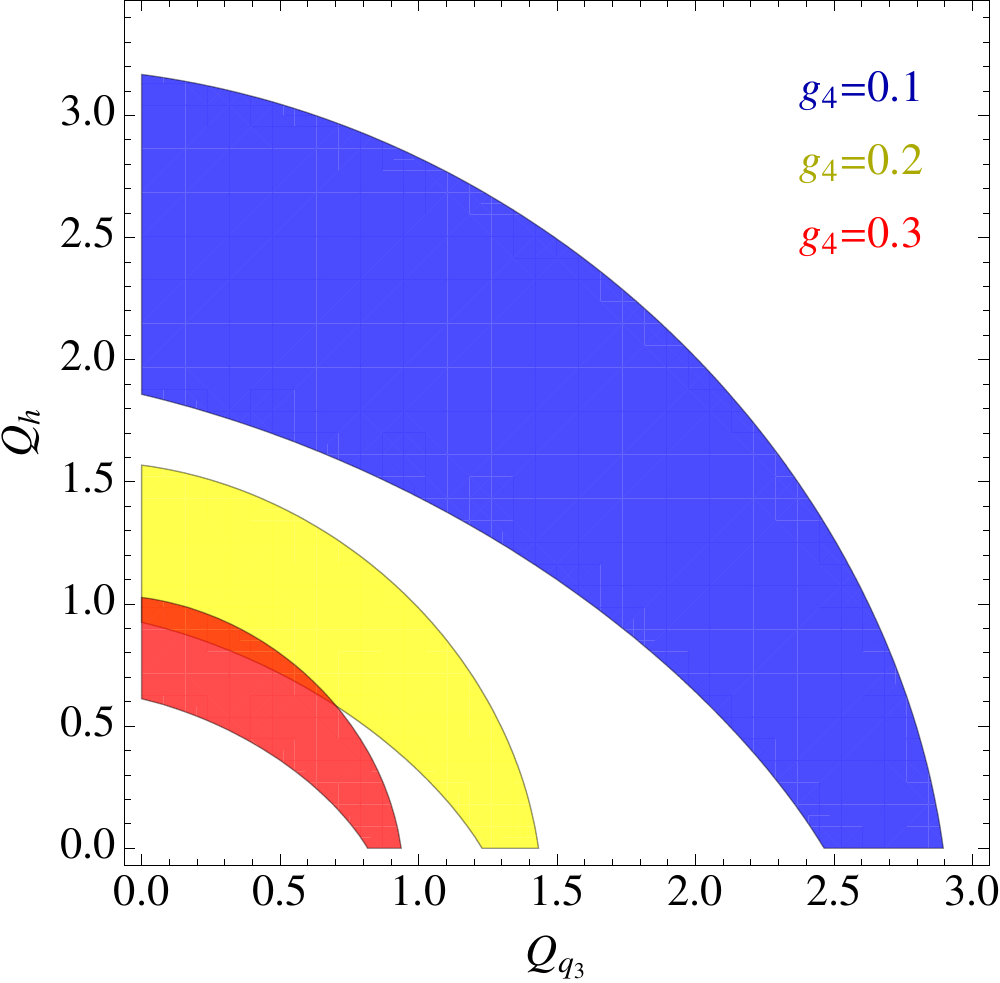}
\includegraphics[width=0.42\linewidth]{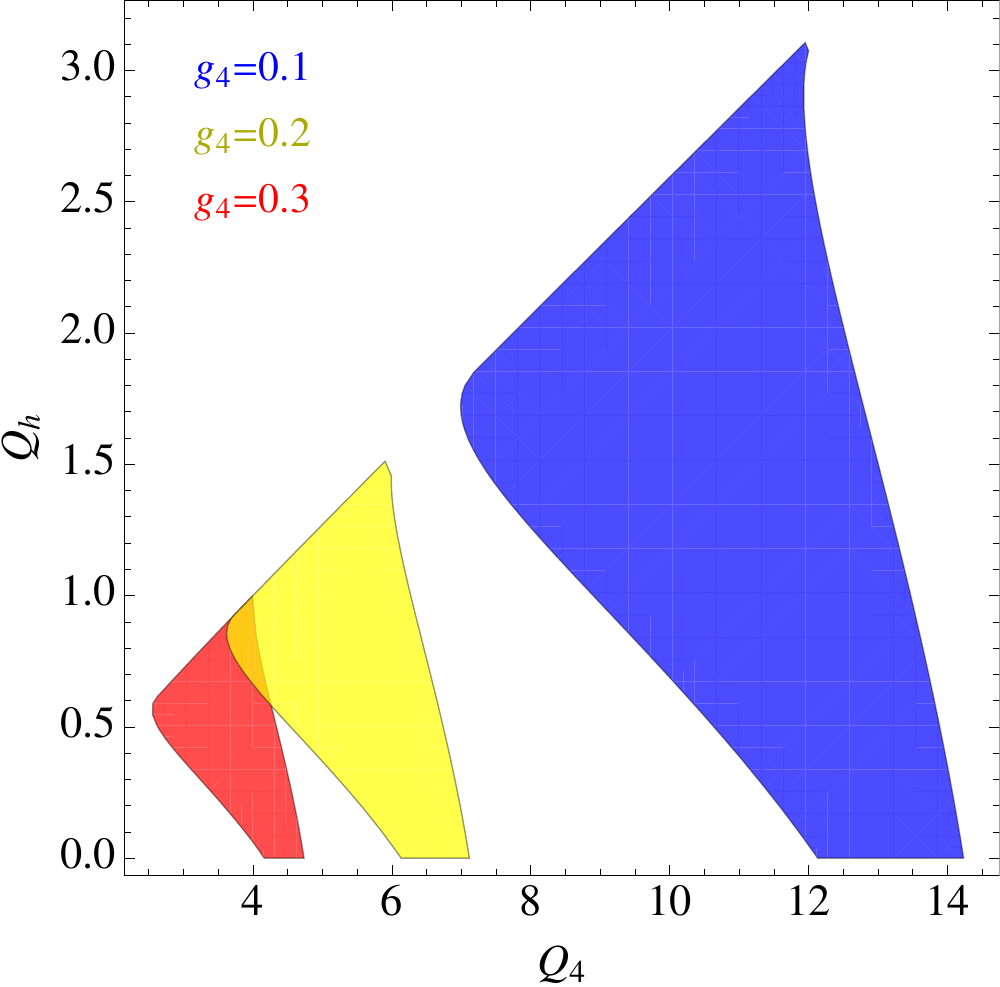}
\caption{Charge assignments consistent with Higgs potential stability and perturbativity for  $g_4=0.1$ (blue region), $g_4=0.2$ (yellow region) and $g_4=0.3$ (red region) at $M_{Z'}=m_t$.}
\label{region-plot}
\end{figure}

For a leptophobic $Z'$, the strong LEP and LHC  bounds on
Z$'$ do not apply, allowing for $M_{Z'} \sim 200$ GeV. 
In this case, the U(1)$'$ affects the running of the couplings already at the electroweak scale. The shape of the allowed regions remains the same as 
in the $M_{Z'} = 3$ TeV case (Fig.~\ref{region-plot}), yet there are important quantitative differences. In particular, the $Q_h=0$ assignment becomes allowed. In that case, the stabilizing effect is due entirely to the reduction
of the top Yukawa coupling. Values of $Q_{q_3}$ between 1 and 3, 
depending on the gauge coupling,  are sufficient to stabilize the scalar potential. The Z$'$ mass plays a crucial role here: the top Yukawa coupling
has its strongest effect on $\lambda_h$ at low energies and reducing $y_t$
in this energy range brings about the desired result. As we have established in the previous subsection, this effect cannot be achieved for a heavy Z$'$: 
compensating the shorter running range by
increasing the charge or $g_4$ carries the coupling into a non--perturbative territory.

\begin{figure}[h!]
\centering
\includegraphics[width=0.493\linewidth]{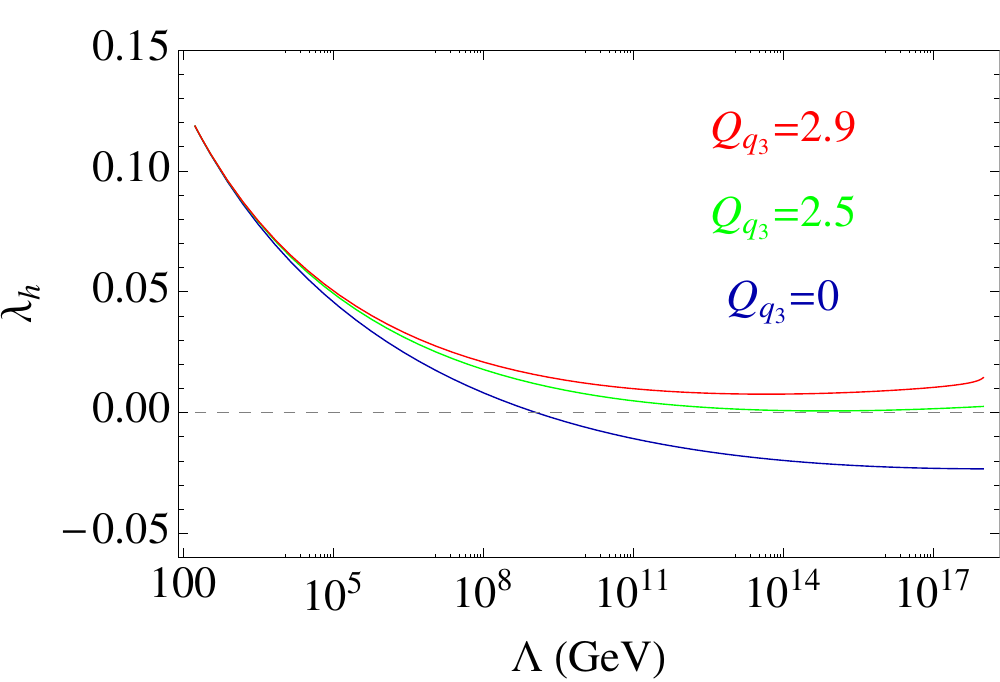}
\includegraphics[width=0.493\linewidth]{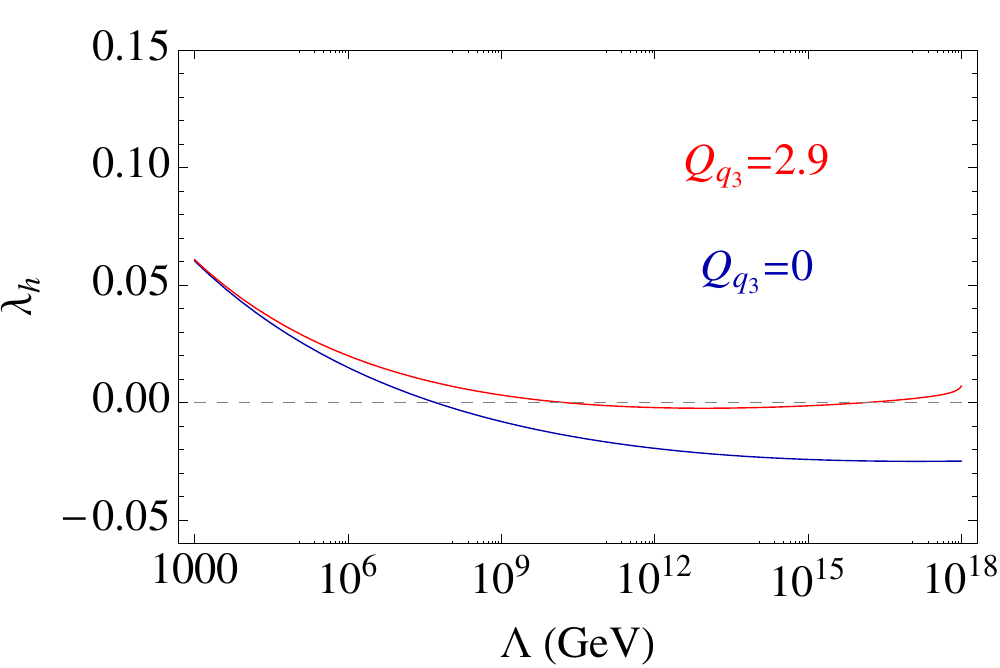}
\caption{Evolution of $\lambda_h$ with energy scale $\Lambda$ for $M_{Z'}=m_t$ (left)
and $M_{Z'}=3$ TeV (right) in the Higgsophobic case. The blue (lower) curve corresponds to the SM, the green (middle) curve corresponds to the minimal $Q_{q_3}$ 
which achieves the Higgs potential stabilization, and the red (upper) curve
corresponds to the maximal allowed $Q_{q_3}$ consistent with perturbativity.
Here $Q_h=0$, $g_4(M_{Z'})=0.1$. 
}
\label{non-zero-g5-B-L}
\end{figure}

The effect of the top quark charge on the evolution of $\lambda_h$ is shown in
Figure \ref{non-zero-g5-B-L}. In the left panel, the Z$'$ has an electroweak mass $\sim m_t$, while
in the right panel $M_{Z'}=3$ TeV.  
Taking $Q_h=0$, the Higgs self--coupling turns
negative at around $10^9$ GeV if the top quark does not couple to the Z$'$.
Increasing $Q_{q_3}$ to about 2.5 with $g_4(m_t)=0.1$ makes the Higgs potential stable up
to the Planck scale, whereas a further increase above 2.9 makes
the system non--perturbative. For the heavy Z$'$, perturbativity is violated
before stability is achieved. At the critical value $Q_{q_3}=2.9$, the Higgs
self--coupling is positive at the Planck scale, yet it is negative at intermediate scales indicating the existence of a deep minimum at
these field values.

We therefore conclude that the Higgs potential can be stabilized even by a $Higgsophobic$ U(1), if the corresponding Z$'$ is light enough. The $B-L$ 
symmetry however does not fall into this category since Z$_{B-L}'$ 
is constrained to  be rather heavy.

\subsection{Effect of the kinetic mixing}

Generally, an extra U(1) mixes with the SM hypercharge.
That implies, among other things,  that even though the charges for the SM fields are zero, the Z$'$ may still couple to SM matter if the kinetic mixing parameter $\epsilon$ is non--zero. 
In this case, $\epsilon$ is constrained to be of order $10^{-2}$ 
for  $M_{Z'}$ of the order of the electroweak scale \cite{Hook:2010tw}. 
The resulting effect on the evolution of $\lambda_h$ is negligible
since for $Q_h=0$ the kinetic mixing contribution to $\beta_{\lambda_h}$
is proportional to $\epsilon^2$  according to Eq.~\ref{RG1}.  

\begin{figure}[h!]
\centering
\includegraphics[width=0.435\linewidth]{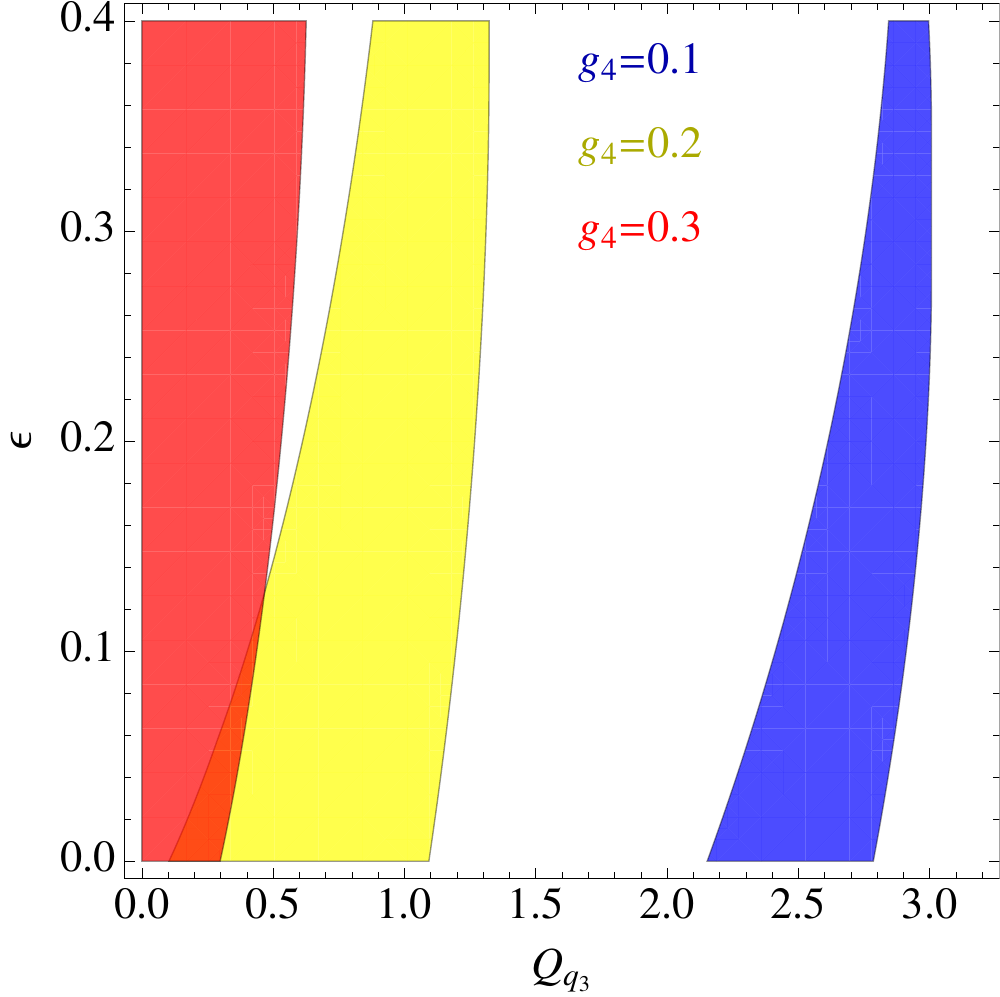}
\caption{Effect of the kinetic mixing $\epsilon$ on the allowed parameter space  for  $g_4=0.1$ (blue region), $g_4=0.2$ (yellow region) and $g_4=0.3$ (red region) at $M_{Z'}=3$ TeV and $Q_h=1$.}
\label{region-plot-general-epsilon}
\end{figure}

The bound on $\epsilon$ relaxes significantly for a heavier Z$'$ 
allowing $\epsilon  \sim {\cal O} (10^{-1})-1 $ at $M_{Z'}~\sim 2-3$ TeV
(see e.g. \cite{Cline:2014dwa}). Of course, $\epsilon ~\sim 1$ 
can simply be reinterpreted as a different U(1)$'$ charge assignment 
with order one charges proportional to the hypercharge (see Eq.~(\ref{D})).
On the other hand, $\epsilon ~\sim 0.1$ is still meaningful as it corresponds
to small charges which otherwise would be unnatural.
For such values of the kinetic mixing, the effect on $\lambda_h$ is
substantial only if $\beta_{\lambda_h}$ contains  linear terms  in $\epsilon$,
that is if $Q_h \not= 0$.
Figure \ref{region-plot-general-epsilon} shows the effect of $\epsilon$ 
on the allowed parameter regions at $Q_h=1$. The kinetic mixing tends to
decrease $\lambda_h$ due to the  $\epsilon$--linear terms in 
$\beta_{\lambda_h}$ of Eq.~\ref{RG1}, which shifts the allowed regions
to somewhat larger charges. 
This behaviour is reversed for negative $\epsilon Q_h$.

 We have also investigated the possibility of stabilizing the Higgs potential 
 entirely due to the kinetic mixing of U(1)$_{B-L}$ and U(1)$_Y$, which belongs to the $Q_h=0$ category. We find that the required $\epsilon$ 
 is too large ($\sim {\cal O}(0.5)$) for the $B-L$ interpretation of the charge assignment to make sense. In that case, however, the scalar contribution from $\lambda_{hs}$ can be efficient in stabilizing the
 Higgs potential.

\section{Conclusion}

We have analyzed the possibility of stabilizing the Higgs potential with  a  Z$'$ boson. We find that a generic Z$'$ improves stability of the potential
in two ways: it increases the beta function of the Higgs quartic coupling
directly and reduces the top quark Yukawa coupling, which also has a positive effect on $\lambda_h$. The Higgs and top quark U(1)$'$ charges play the most important
role in this mechanism. The stabilization is achieved for order one charges and the
gauge coupling $g_4$ of the electroweak size. In case of a light Z$'$, 
$M_{Z'} \sim m_t$, the Higgs potential can be stabilized even if the Z$'$ does not 
couple to the Higgs, i.e. entirely through a reduction of the top Yukawa coupling.
A heavier Z$'$ in the multi--TeV range necessitates a direct coupling to the Higgs 
boson to achieve the same effect.

We have also analyzed the effect of a kinetic mixing term between the Z$'$ and the hypercharge. We find that the mixing parameter $\epsilon$ of order $10^{-1}$ can have a tangible effect on the allowed parameter space.
 
{\bf Acknowledgements.} This work was financially supported by the Academy of Finland project ``The Higgs Boson and the Cosmos'' and  project 267842.

\end{document}